\documentstyle[aps,preprint,tighten,epsfig]{revtex}
\begin{document}
\draft
\title
{
Momentum  Sum Rules in QCD for
a Photon Target}
\author{L.~L.~Frankfurt\cite{em1,leave} and
E.~G.~Gurvich\cite{em2}}

\address{ School of Physics and Astronomy \\
Raymond and Beverly Sackler Faculty of Exact Sciences\\
Tel Aviv University, Ramat Aviv 69978, Israel}
\date{\today}
\maketitle
\begin{abstract}

  We deduce   momentum   sum rules for the parton     
structure functions of a photon target.
Non-perturbative QCD contribution to the momentum sum rules
follows from conservation of the energy--momentum tensor and
it is calculated through the
hadronic part of the photon vacuum polarization
operator. The contribution of the unresolved photon
unambiguously follows from gauge invariance, renormalizability
and asymptotic freedom in QCD.
We also compare available parametrizations of  parton distribution in 
a photon with the deduced sum rules.

\end{abstract}

\narrowtext
      The aim of  this study is to
deduce the momentum sum rule (MSR)
for  the case of parton distributions (PD) in a photon target.
A naive expectation based on  the parton model approximation is 
that the integral over $x$ with the weight $x$ 
from the sum of parton distributions  
of a (quasi)real photon   target   
should be 
equal to the
probability of a target photon to be in a hadron configuration,
 $Z^{-1}(\gamma \to \rm hadrons)$. This
expectation is invalid in QCD and QED. The nontrivial contribution
to the photon wave function 
due to configurations of quarks
with large relative transverse momenta
(the unresolved photon),
can not
be described in terms of the parton model.

Throughout this  Letter we denote  the four momenta of the probing
virtual  photon and of the target photon by $q$ and $p$ respectively.
The mass squared
of the probing  photon (target photon)
is  $q^2=-Q^2 $ \hbox{$(p^2 = -P^2)$}. With $\nu = 2qp$\/,
the Bjorken variable $x=Q^2/\nu $.
We define the parton distribution functions of flavour $f$ as
$q^{\rm T (L)}_f = {F_{2\  \gamma(P^2)}^{f \rm T(L)} (x,Q^2)
\over e_f^2 x} $, where
$F_{2\  \gamma(P^2)}^{f \rm T(L)} (x,Q^2)$ is the
conventional structure function  for
a target with definite helicity T,L
and includes the contribution of the unresolved photon.$e_f$ is the 
electric charge of flavour $f$.
Within the parton model these structure
functions
describe  the flavour
singlet quark distributions
$S^{\rm T(L)}_{\gamma}(x,Q^2,P^2)
\equiv \sum_{f = 1}^{N_f} (q^{\rm T(L)}_{f} + \bar{q}^{\rm T(L)}_{f})$.
The distribution of gluons is then
 $G^{\rm T(L)}_{\gamma }(x,Q^2,P^2) =
\sum_f G^{f \rm T(L)}_{\gamma }(x,Q^2,P^2)$,
where we
account for the fact that the
gluon distribution may differ for different flavours.

Let us  generalize the MSR for a virtual photon
target in  QED with one lepton flavour.
In the lowest order of perturbation theory in $\alpha_{\rm em}$,
the photon structure functions are given by the imaginary part
of the sum of the box diagrams of Fig.1,  for both
the real and
the virtual target photons.
We work in  the Bjorken limit $Q^2 \gg \mu^2,P^2$;
$x$ is fixed, with
$\mu$ the mass of the lepton.
We start our considerations from the case of a longitudinally polarized
photon target.
The polarization vector of  the longitudinal
polarization is
$e_L={(p_3,{\bf 0},p_0)/ \sqrt{p_3^2-p_0^2}}$,
where $(p_0,0,0,p_3)$ is four momentum of target photon.
We denote by axis 3 as the direction of the momentum of the target photon.
Using conservation of the electromagnetic current
$p_{\mu} J_{\mu}^{em}=0$
we obtain,
\begin{equation}
e^{\rm L}_{\mu} e^{\rm L}_{\nu} M_{\mu \nu}^{\alpha \beta}=
P^2 {M_{00}^{\alpha \beta} \over p_3^2}\, ,
\label{5c}
\end{equation}
where $ M_{\mu \nu}^{\alpha \beta}$ is the amplitude of $\gamma^*(q)
\gamma^*(p)$ scattering. The upper indices correspond to the
polarization of
the
projectile photon $\gamma^*(q)$ and bottom indices correspond to
polarization of target photon $\gamma^*(p)$.
The sum of Feynman diagrams for ${M_{00}^{\alpha \beta}}$ is
superconvergent.
It is thus legitimate to apply the traditional
machinery of the parton model, the Wilson Operator Product Expansion (OPE), to
${M_{00}^{\alpha \beta} /  p_3^2}$
and to use the conservation of the energy--momentum tensor
to deduce the MSR.
We do not  decompose the
scattering amplitude into
independent Lorentz invariant structures,
since it  contains
more independent invariant functions than for  a nucleon target,
as  we do not sum
over the polarization of the target photon.
We derive the sum rule for
the structure function
of the photon target,
\begin{equation}
\int_{0}^{1} x \mbox{d}x {\nu \over p^2_3}
\left({{P^2 M_{00}^{33}\over p_3^2}}\right)=
\int_{0}^{1}\mbox{d}x{{xS^{\rm L}(x,Q^2,P^2)}}
=-{{\rm d}\pi(P^2) \over {\rm d}\ln P^2 }\, .
\label{6c}
\end{equation}
Here  $\pi(p^2)$ is directly expressed through the polarization operator
$\Pi_{\mu\nu}$
of a target photon $\gamma^*(P^2)$,
\begin{equation}
\Pi_{\mu \nu}=(p^2\delta_{\mu \nu} -p_{\mu}p_{\nu})\pi(p^2).
\label{7c}
\end{equation}
The relationship between the amplitude  $ M_{\mu \nu}^{\alpha \beta}$
and the structure function follows from the calculation of
Feynman diagrams
in the reference frame where
the momentum of the target $p\rightarrow \infty $
but $q_0$ is small \cite{Feynman}. We choose to work
in the center of mass system
of the projectile electron and the target photon $\gamma^*(P^2)$.
In this frame
\hbox{$q=((\nu+Q^2)/4|{\bf p}|,{\bf q_t},(\nu-Q^2)/4|{\bf p}|)$}.
Thus the MSR for the electron-positron
distributions  within the longitudinally
polarized photon target is,
\begin{equation}
\zeta_{\rm L}^{-1} \equiv \int_{0}^{1}\mbox{d}x {xS_{{\rm e}}^{\rm L}
  (x,Q^2,P^2)} = -{{\rm d} \pi(P^2)\over {\rm d}\ln(P^2)} .
\label{8c}
\end{equation}

Let's now turn  to the case of a transversely polarized virtual photon.
A na\"\i ve application of
the impulse approximation leads to
the conventional MSR 
where in difference from Eq.(\ref{8c}) 
the right hand side is given by  the normalization
of the photon wave function,
{\it i.e.}\/ by the probability of the target photon to be
in an ${\rm e}^+{\rm e}^-$ configuration 
$Z^{-1}_{em}$.
The renormalization ``constant'' 
 $Z^{-1}_{\rm T} \equiv Z^{-1}_{em} (P^2) =
{{\rm d}\over {\rm d} P^2} [P^2 \pi(P^2)]$
of the virtual photon $\gamma^*(P^2)$
is logarithmically ultraviolet divergent.
Thus the parton model
is applicable for
the calculation of the difference between PD in target photons
with different virtualities (which is ultraviolet finite)
but not for PD themselves,
\begin{eqnarray}
  \label{delta}
\nonumber \zeta_{\rm T}^{-1} & \equiv &
\int_0^1 {\rm d} x x [S^{\rm T}_e(x,Q^2,0) -
S^{\rm T}_e(x,Q^2,P^2) ]\\
& = &
   Z^{-1}_{\rm T, em}(0) - Z^{-1}_{T,em} (P^2)
=
{{\rm d}\over {\rm d} P^2} \{P^2 [\pi(0)-\pi(P^2)]\} .
\end{eqnarray}

For the sake of generalization to QCD it is instructive to explain
the problem in terms of the parton model description.
The parton wave function
of the photon $\Psi_{\gamma\to e \bar{e}} (x_1,x_2,k_t)$ \cite{BrLep}
is given by the electromagnetic transition
 $\gamma^*\rightarrow {\rm e}\bar{\rm  e}$
which includes the energy  denominator. Here $x_i(k_t)$ is
the light--cone fraction
of the photon momentum (transverse momentum) carried by
the electron and the positron. For large $k_t^2$ the
wave function  is \hbox{$|\Psi |^2 \sim 1/k_t^2$}
and therefore $\int |\Psi |^2\mbox{d}^2k_t$
diverges logarithmically.
If  $k_t^2 \gtrsim Q^2$, the probing photon interacts coherently with
the $e\bar{e}$ pair in a target photon .
There is a destructive interference  between
the diagrams for a structure
function where the virtual photon interacts with one parton (parton model
contribution)  and the diagrams where  the virtual photon interacts with
both constituents of the photon. The cancellation between these
diagrams, the charge screening phenomenon, effectively cuts the
integration over $k_t^2$ at a value  $\sim Q^2$. So a correct treatment
of the contribution of  the unresolved photon leads to a finite
value of the matrix element,
but the impulse approximation and therefore the momentum sum rule are  lost.
At the same time the unresolved photon contribution
in Eq.(\ref{delta})
is cancelled since it is a high
transverse momentum contribution and it does not depend
on virtuality
of the  target photon
(within the power accuracy  over $P^2\over Q^2$) .

It is easy to demonstrate that the above reasoning agrees with
results
of the most complete calculation of box diagrams
with virtual photons~\cite{Budnev}.

To generalize above results to QCD 
let's consider now the case of an unpolarized target photon.
The standard definition of the photon structure function
$F_2 = {1 \over 2}\sum_\lambda
M_{\mu\nu}^{33} e^\mu_\lambda e^\nu_\lambda\nu / p_3^2$
 (see for example the first paper of
Ref.\cite{VirPhot}) leads to the definition of PD for unpolarized
virtual photon as follows:
\begin{equation}
\label{np}
  S_{\gamma^*(P^2)}(Q^2,P^2,x) =
 S^{\rm T}{\gamma^*(P^2)} (Q^2,P^2,x) - 
{1 \over 2} S^{\rm L}{\gamma^*(P^2)}(Q^2,P^2,x).
\end{equation}
And then the MSR for the case of parton distributions in
the unpolarized virtual
photon has the  form,
\begin{equation}
\int_0^1 x [ S_{\gamma^*(P^2)}(x,Q^2,P^2)+ 
G_{\gamma^*(P^2)}(x,Q^2,P^2) ] {\rm d}x =
\zeta_{\rm T} (Q^2,P^2) -{1 \over 2} \zeta_{\rm L} (Q^2,P^2),
  \label{NPol}
\end{equation}
where $\zeta_{\rm T}$ and $\zeta_{\rm L}$ are given in QED by
Eqs.(\ref{8c},\ref{delta}). Let us calculate $\zeta_{\rm T}$ and $\zeta_{\rm L}$
in QCD. 
The important difference between QED and QCD is that in QCD
the constituents
are quarks and gluons and that the box diagrams do not account
for the full structure
functions of a (quasi)real photon.

The derivation of the MSR in
 QCD consists of two steps. The first step is to apply OPE, 
the QCD improved parton model,
to the
difference of  the structure functions of target photons with
virtualities $P ^2$ and $K^2$. 
In this difference the contribution of
the unresolved photon into MSR is
canceled out as in QED.
This cancellation is 
evident in terms of Feynman diagrams since
the contribution of the unresolved photon corresponds to virtualities
of partons  $\sim Q^2 \gg P^2,K^2$.  Therefore the  contribution
of unresolved photon is 
independent on the virtuality of target photon.  As a result the validity 
of the QCD improved 
parton model approximation in calculating the MSR for the
difference of structure functions of  photon targets with
different  
virtualities  can be justified.  For the leading twist 
contribution the MSR for the difference of structure functions has the same 
form as in QED (see Eqs.(\ref{8c},\ref{delta})), 
\widetext 
\begin{eqnarray} 
\label{QCD-L} && \zeta_{\rm L}^{-1}(Q^2,P^2) =  -{{\rm d}\pi_{\rm had} 
(P^2) \over {\rm d}\ln P^2 } \\ 
\label{QCD-R} &&  \zeta^{-1}_{\rm T} (Q^2,P^2) - \zeta^{-1}_{\rm T}(Q^2,K^2) 
= {{\rm d} \over {\rm d} P^2} [ P^2 \pi_{\rm had}(P^2)]  - 
{ {\rm d}\over {\rm d} K^2} [ K^2 \pi_{\rm had}(K^2)].  
\end{eqnarray} 
\narrowtext
Here $\zeta^{-1}_{\rm T (L)}$ is the normalization of parton distributions
for  transversely  (longitudinally) polarized photons
and includes now the quark, antiquark and gluon
contributions,
\begin{eqnarray}
\nonumber
&&\zeta_{\rm L(T)}^{-1}(Q^2,P^2) =\\
  \label{Z}
&&\int_0^1 {\rm d}x x
[ S^{\rm L(T)}_{\gamma} (x,Q^2,P^2)
+ G^{\rm L(T)}_{\gamma} (x,Q^2,P^2)].
\label{defz}
\end{eqnarray}
and the $\pi(P^2) \equiv \pi_{\rm had} (P^2)$ is the hadronic
contribution to the renormalized photon vacuum polarization operator.
It follows from the above discussion
that the precision of Eqs.(\ref{QCD-L},
\ref{QCD-R}) is the same as that for the factorization theorem in QCD.

One of the practical applications of Eqs.(\ref{QCD-L}-\ref{defz})
is the possibility to
measure the dependence of the fraction of the photon momentum carried by gluons
on the virtuality of the photon.
This would be feasible if the
momentum carried by quarks of the virtual photon
could be  measured experimentally.

Thus we deduce
the momentum sum rule for the photon structure function
as follows:
\begin{equation}
\zeta^{-1}_{\rm T} (Q^2,P^2) = \zeta^{-1}_{\rm T}(Q^2,K^2) 
+ {{\rm d} \over {\rm d} P^2} [ P^2 \pi_{\rm had}(P^2)]  - 
{ {\rm d}\over {\rm d} K^2} [ K^2 \pi_{\rm had}(K^2)].  
\label{real}
\end{equation}
where
\begin{equation}
\pi_{\rm had}(0) - \pi_{\rm had}(K^2)= (1/4\pi ^2 \alpha_{\rm em})
K^2\int_{0}^{\infty}\mbox{d}s \sigma_h(s) /(s+K^2),
\label{Pi}
\end{equation}
and $\sigma_h(s) \equiv \sigma_{\rm e^+e^- \to hadrons} (s)$ with $s$
the c.m.s. energy square.

The second step in the derivation of the MSR
is to choose large $K^2$ such as \hbox{$\Lambda^2_{\rm QCD} \ll K^2
\ll Q^2$} where it is legitimate (see Ref.\cite{VirPhot}) to apply
 perturbative
QCD and asymptotic freedom to calculate $\zeta^{-1}_\gamma(Q^2,K^2)$.
It follows from the renormalizability of QCD that  $\pi(K^2)$
can be represented at large $K^2$
as an asymptotic series in powers of $\alpha_s(K^2)$,
\begin{eqnarray}
  \label{PiPert}
  \pi(0) - \pi(K^2) =
\sum_{r=0} [\alpha_s(K^2)]^r [c_r \ln {K^2 \over \Lambda^2_{\rm QCD}}
+  d_r ]\,.
\end{eqnarray}
Here $c_r$ and $d_r$ are some numerical coefficients.
It follows from Eq.(\ref{real}) that the same decomposition is valid for
$\zeta^{-1}_{\rm T}(Q^2,K^2)$ since the l.h.s. of Eq.(\ref{real}) is
independent of $K^2$. Thus to calculate the r.h.s. of  Eq.(\ref{real})
at large $Q^2$
it is sufficient to keep in the polarization operator
 $\pi(K^2)$ and in $\zeta^{-1}_{\rm T}(Q^2,K^2)$  only terms of zero order in
 $\alpha_s(K^2)$ . Other terms cancel out in the r.h.s.
of Eq.(\ref{real}).
But the lowest order term in $\alpha_s(K^2)$
for $\zeta^{-1}_{\rm T}(Q^2,K^2)$ is given by the sum of QED box diagrams
multiplied by the factor $N_c \sum_f e_f^2$,
\begin{equation}
\zeta^{-1}_{\rm T}(Q^2,K^2)=N_c \sum_f e_f^2{\alpha_{\rm em} \over 3\pi}
\left[ \ln {Q^2 \over K^2}- {1 \over 12} \right].
\label{boxv}
\end{equation}
It is easy to check that the term $\ln(1/K^2)$ in Eq.(\ref{boxv}) is cancelled
on the r.h.s. of   Eq.(\ref{real}) with the corresponding term in
$\pi(K^2)$.

For the practical purposes,  
Eqs.(\ref{real}--\ref{boxv})
can be simplified since  
for the production of each flavour $f$ one can find such a value $s_0(f)$
that for  
$s>s_0$ 
the contribution of flavour $f$ in
$\sigma_{\rm h}(s)$ 
is given by quark loops without hard QCD radiative corrections.
For $Q^2 \gg s_0(F)$
Eq.(\ref{real}) has the form,
 \begin{eqnarray}
   \nonumber
 &&\zeta^{-1}_{\rm T}(Q^2,P^2)=N_c \sum_{f=1}^F e_f^2{\alpha_{\rm em} 
\over 3\pi}
\left[ \ln {Q^2 \over s_0(F)+P^2}- {1 \over 12}\right. \\
\label{MSRG}
&&+\left.{s_0(F) \over s_0(F) + P^2} \right]
+ (1/4\pi ^2 \alpha_{\rm em}) \int_{0}^{s_0(F)}\mbox{d}s
s^2\sigma_h(s) /(s+P^2)^2.
 \end{eqnarray}

In the above derivation we ignored the threshold
effects related to heavy flavour production.
So our formulae are applicable for  the production of
flavours with masses $M_q^2 \ll Q^2 $.

  Our final result is given  by the Eq.(\ref{NPol}),
where $\zeta^{-1}_{\rm L}$ is given by Eq.(\ref{QCD-L}) 
and $\zeta^{-1}_{\rm T}$ is 
given by Eq.(\ref{MSRG}).
In the approximation of 
flavor $SU(3)$ 
symmetry for the parton distributions in a
photon we can deduce the MSR for the nonsinglet structure
function which is expected to be valid for $P^2 \ll 4M_c^2$,
\begin{eqnarray}
\nonumber
&&  \int_0^1 x \left[ u(x,Q^2,P^2) - d(x,Q^2,P^2) ] {\rm d}x  =
[\zeta^{-1}_{\rm T}(Q^2,P^2)\right.\\
  \label{NS}
&&- {1 \over 2}\zeta^{-1}_{\rm L}(Q^2,P^2) -
\zeta^{-1}_{\rm T,c}(Q^2,P^2)]
{(e_u^2-e_d^2) \over \sum_{u,d,s}e_f^2}.
\end{eqnarray}
By definition $\zeta^{-1}_{\rm T,c}(Q^2,P^2)$ is the contribution
of the charm quark into the
normalization of the structure function calculated
through box diagrams. We use the fact that only
the term $\sim  \ln (Q^2/M_c^2)$
is important and that  this term is dominated by the box diagrams.

It is of interest to  compare the sum rule in Eq.(\ref{MSRG})
with some available
parametrizations of parton distributions in a real photon.
The results are shown in Fig.2.
To estimate  $\zeta^{-1}_{\rm T}(Q^2,0)$ we use the
parametrization of $\sigma_{h} (s)$ from
Ref.\cite{Vys}. It accounts for the production of low mass hadron states and
describes $\sigma_{h}(s)$ at large $s$ in terms of
the parton model contributions with the first order QCD corrections.
(Use of the available experimental
parametrization of  $\pi_{\rm had} (p^2)$ from \cite{Burkh} leads to similar
results).
For the self--consistency of Eq.(\ref{MSRG}) we
neglect the  hard QCD corrections  to the expression
for   $\sigma_{h} (s)$ in the parametrization of \cite{Vys}.
In the calculations of structure functions and
photon vacuum polarization function we also neglect
heavy quark (b,t) effects .

It follows
from Fig.2 that the QCD sum rule,
Eq.(\ref{MSRG}), predicts 
smaller  second moment of the sum of parton  structure functions of a photon
as compared to  the existing  parametrizations \cite{LAC,DG,GRV,Hag}.
To visualize
the difference it is convenient to represent  the sum rule  
for $Q^2 < 10{\rm GeV}^2$ in the form,
\begin{eqnarray*}
 {1 \over \alpha_{em}}
\int_0^1 x [ S_{\gamma^*(P^2=0)}+ G_{\gamma^*(P^2=0)} ]{\rm d}x\\
= N_c \sum_f e_f^2{1 \over 3\pi}
 \ln {Q^2 \over 4{\rm GeV}^2} + c.
\end{eqnarray*}
The parametrizations of  Ref.\cite{LAC}, of Ref.\cite{GRV} and
of Refs.\cite{DG,Hag}
correspond to
 $c \approx 4, 2, 1.5$ respectively, while the
QCD sum rule deduced in this  Letter
corresponds to  $c \approx 1$.

The authors are indebted to H.~Abramowicz and A.~Levy
who drew our attention to the problem of the MSR for the photon target.
One of the authors (L.F.) is indebted to
 D.~Soper for the explanation of specific features
of OPE as applied to $\gamma^*\gamma$
scattering.
This work is partially supported by the Israeli Academy of Science
and GIF grant No.I 0299-095.07/93.

\begin{figure}\begin{center}
\mbox{\epsfig{file=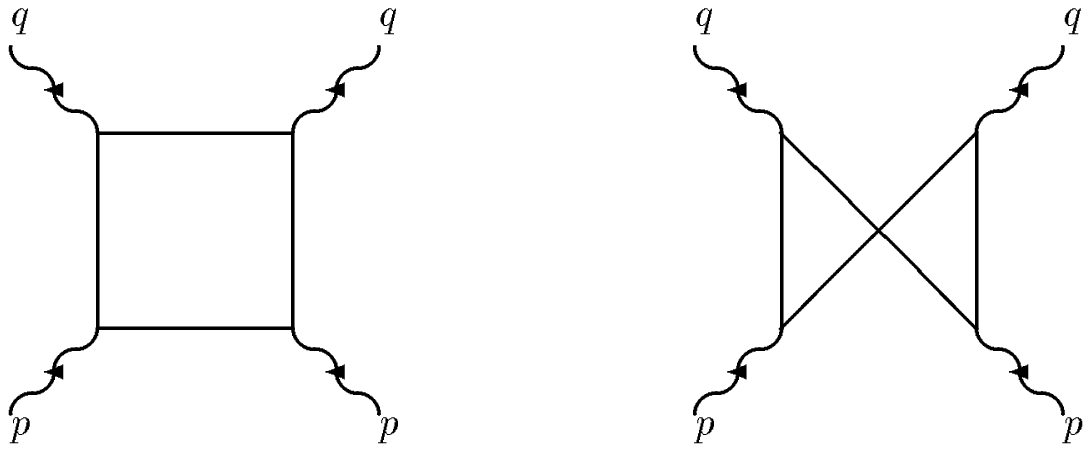,width=3.375in}}
\end{center}
\label{box}
\caption{QED box diagram}
\end{figure}
\begin{figure}
  \begin{center}

   \mbox{\epsfig{file=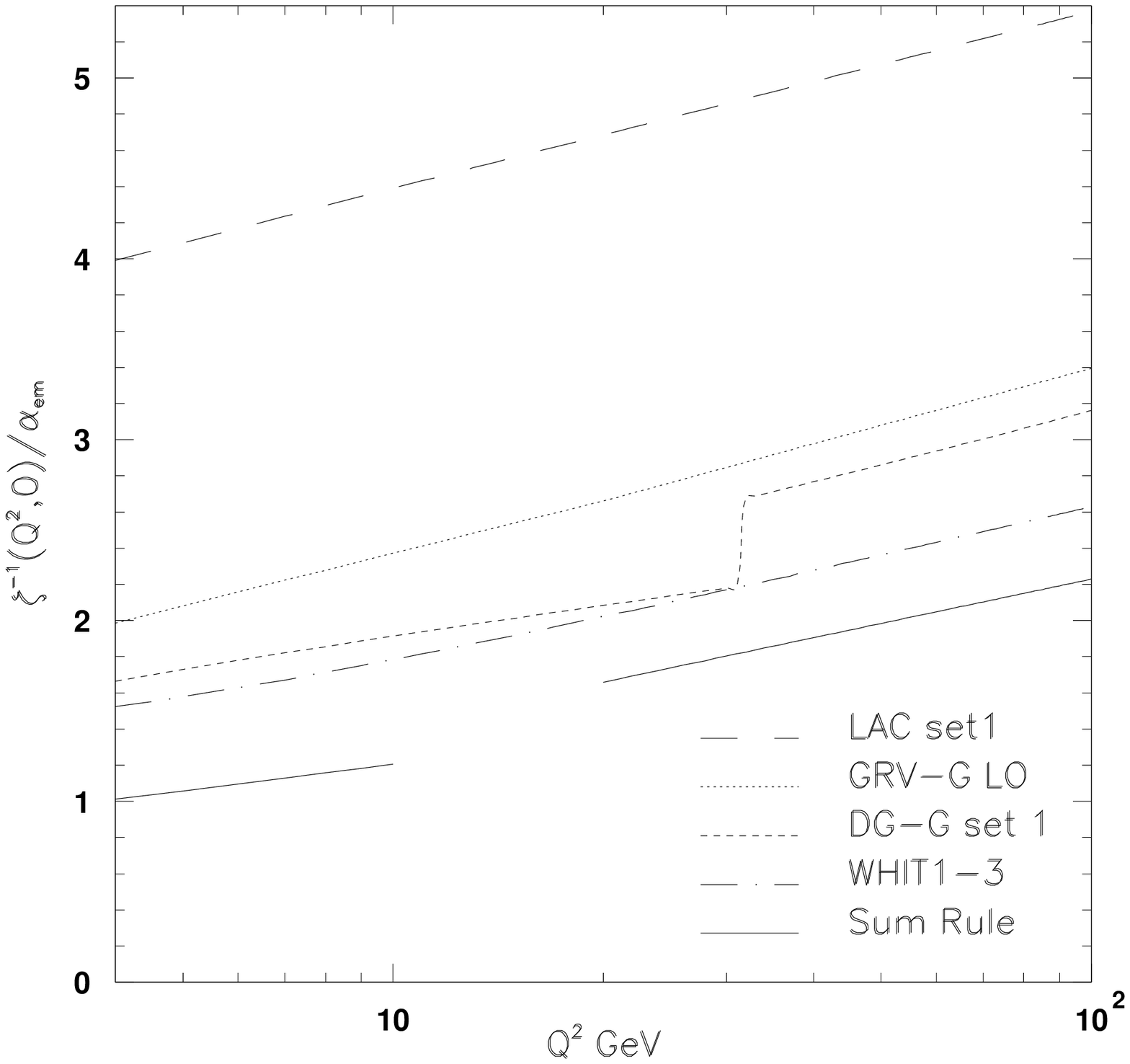,width=3.375in}}
  \end{center}
\end{figure}
 FIG.2.~Compaison of the momentum sum rule for the real photon
target with 
parametrizations of the photon PD,
LAC from \cite{LAC}, GRV-G LO from \cite{GRV},
DG-G, set 1 from \cite{DG} and WHIT1-3 from \cite{Hag}.
The full line is the  MSR prediction for $N_f = 3$,
 $Q^2 \le 10{\rm GeV}^2$,  
and for
$N_f = 4$ , $Q^2 \ge 20{\rm GeV}^2$.
\end{document}